\documentclass[twocolumn,a4paper,groupedaddress,showpacs,floatfix,aps,10pt,longbibliography,nobibnotes,prxquantum]{revtex4-2}

\usepackage{graphicx,amsmath,amssymb,amsfonts,dsfont,subfigure,xcolor}

\newcommand{\mc}[1]{\ensuremath{\mathcal{#1}}}

\usepackage{color}

\makeatletter
\newcommand{\fmarki}{*}
\newcommand{\fmarkii}{*\ensuremath{\dagger}}
\newcommand{\fmarkiii}{\ensuremath{\dagger\dagger}}

\def\@fnsymbol#1{{\ifcase#1\or \fmarki\or \fmarkii\or \fmarkiii \else\@ctrerr\fi}}
\makeatother
\begin{document}

\title{Imaging a chain of strongly correlated Rydberg excitations enabled by F\"{o}rster-resonance-enhanced interaction}

\author{Jinjin Du${}^{1,2}$}
\thanks{These two authors contributed equally.}
\author{Thibault Vogt${}^{1,3,5}$}
\thanks{These two authors contributed equally.}
\author{Ningxuan Zheng${}^{1}$}
\author{Wenhui Li${}^{1,4,6}$}

\affiliation{Centre for Quantum Technologies, National University of Singapore, 3 Science Drive 2, Singapore 117543${}^1$}
\affiliation{Quantum Machines Unit, Okinawa Institute of Science and Technology Graduate University, Onna, Okinawa, 904-0495, Japan${}^2$}
\affiliation{School of Physics and Astronomy, Sun Yat-sen University, Zhuhai, 519082, China${}^3$}
\affiliation{State Key Laboratory of Quantum Optics and Quantum Optics Devices and Institute of Opto-Electronics, Shanxi University, Taiyuan 030006, China${}^4$}
\affiliation{ttvogt@mail.sysu.edu.cn${}^5$}
\affiliation{liwenhui@sxu.edu.cn${}^6$}

%\author[1,2,$\dag$]{Jinjin Du}
%\author[1,3,$\dag$,5]{Thibault Vogt}
%\author[1]{Ningxuan Zheng}
%\author[1,4,*]{Wenhui Li}

%\affil[1]{Centre for Quantum Technologies, National University of Singapore, 3 Science Drive 2, Singapore 117543}
%\affil[2]{Quantum Machines Unit, Okinawa Institute of Science and Technology Graduate University, Onna, Okinawa, 904-0495, Japan}
%\affil[3]{School of Physics and Astronomy, Sun Yat-sen University, Zhuhai, 519082, China}
%\affil[4]{State Key Laboratory of Quantum Optics and Quantum Optics Devices and Institute of Opto-Electronics, Shanxi University, Taiyuan 030006, China}
%\affil[5]{ttvogt@mail.sysu.edu.cn}
%\affil[$\dag$]{These two authors contributed equally.}
%\affil[*]{liwenhui@sxu.edu.cn}


% 42.50.Gy Effects of atomic coherence on propagation, absorption, and amplification of light; electromagnetically induced transparency and absorption
%
% 42.65.Ky Frequency conversion; harmonic generation, including higher-order harmonic generation
%
% 32.80.Ee Rydberg states
%

\begin{abstract}
Rydberg atoms are currently a very fast advancing quantum platform. For many interesting and demanding applications, including quantum computation, fast detection of a Rydberg excitation or a Rydberg qubit for information readout would be one of the most desirable developments. We demonstrate single-shot and \textit{in situ} absorption imaging of individual Rydberg excitations. This level of resolution is achieved using an electromagnetically induced transparency scheme involving a Rydberg energy level that is highly sensitive to the presence of Rydberg atoms due to F\"{o}rster-resonance-enhanced dipole couplings. Spectroscopic measurements illustrate the existence of the F\"{o}rster resonance and underscore the state-selectivity of the technique. With an imaging exposure time as short as 3 $\mu$s, we successfully resolve linear chains of Rydberg excitations in a one-dimensional configuration. The extracted second-order correlation shows strong anti-bunching due to excitation blockade, and a Fourier analysis reveals the long-range order in the chains of Rydberg excitations. This imaging technique, with minimal destruction, will be of great interest for leveraging ensemble-encoded qubits in quantum computation and quantum simulation applications.
\end{abstract}

\maketitle

\section{Introduction}
\label{introduction}

Highly excited Rydberg atoms have emerged as a prominent physical platform for applications in quantum nonlinear optics~\cite{firstenberg2016nonlinear} and quantum simulation and computation~\cite{saffman:10,browaeys2020}. The essential element enabling the high degree of control and manipulation at the single-particle level in these systems is the strong long-range interaction between Rydberg atoms~\cite{gallagher:ryd}. The resulting excitation blockade effect~\cite{lukin:01,vogt2006dipole,gaetan:09,urban:09} leads to collective and entangled quantum states, which are important resources for quantum applications. Fascinating demonstrations in quantum optics include generating single-photon sources~\cite{dudin2012strongly,peyronel2012quantum}, single-photon switches~\cite{baur2014single}, single-photon transistors~\cite{gorniaczyk2014sptransistor,tiarks2014sptransistor,gorniaczyk2016}, phase gates~\cite{tiarks2019photongate}, and contactless non-linear optics~\cite{busche2017contactless}. More recently, Rydberg-atom arrays, exploring the long-range interaction in ordered configurations, have made remarkable progress in simulating quantum many-body physics and advancing large-scale neutral-atom quantum computation~\cite{browaeys2020,ebadi2021,scholl2021,kim2022,bluvstein2023}. 

So far, the detection of Rydberg atoms has mostly relied on destructive means. For an extended period of time, state-selective field ionization served as the primary method for conducting measurements on Rydberg atoms~\cite{gallagher:ryd}. It has recently been further developed into high-resolution ion microscopy~\cite{schwarzkopf:11,stecker2017high,veit2021pulsed}. When Rydberg excitations are induced in optical lattices or tweezer arrays, they are often detected via fluorescence imaging of ground-state atoms, where the loss of an atom at a particular tweezer or lattice location indicates its excitation to Rydberg states before measurement~\cite{schauss2012observation,labuhn2016tunable}. The seminal work on electromagnetically induced transparency involving Rydberg states (Rydberg EIT)~\cite{mohapatra2007coherent} has laid the foundation and paved the way for the direct optical detection of Rydberg excitations. In Rydberg EIT, long-range interactions can be mapped onto polaritons and detected by retrieving and counting the photons released from the polaritons  ~\cite{dudin2012strongly,peyronel2012quantum,busche2017contactless,tiarks2019photongate}. In the dissipative regime, capitalizing on the interaction-induced scattering effect, Rydberg EIT has recently been employed for nondestructive optical imaging of Rydberg excitations~\cite{olmos2011,gunter2012interaction,gunter2013observing,xu2021rydbergensembels,srakaew2023}. The long-range interaction between one Rydberg excitation and surrounding polaritons is responsible for a sharp change of transmission of EIT, which allows fast \textit{in situ} imaging of the Rydberg excitation in a nondestructive way. This method has been used for imaging transport dynamics of Rydberg excitations, but without reaching single-particle resolution~\cite{gunter2013observing}. The latest development and applications of this detection approach include repeated nondestructive measurements of an ensemble-based Rydberg qubit~\cite{xu2021rydbergensembels}, switching an atomically thin mirror by a single Rydberg atom~\cite{srakaew2023}, as well as fast imaging of individual ions in an atomic ensemble~\cite{du2023}. These experiments have proven that this ensemble-assisted method would be particularly suitable for detecting a Rydberg excitation within a mesoscopic ensemble or a Rydberg ``superatom''~\cite{petrosyan2011electromagnetically,kumlin2023,shao2024}, and is faster and less destructive compared to other means of detecting Rydberg atoms.

In this paper, we demonstrate spatially resolved imaging of a chain of Rydberg excitations in a single shot with a 3 $\mu$s exposure time. We reach this sensitivity by exploring a novel F\"{o}rster-resonance scheme between a 27$F$ state and a 96$S$ state. The former serves as the upper level of the Rydberg EIT for detection, while the latter is the state of Rydberg excitations under investigation. The enabling strong long-range interaction arises from a F\"{o}rster-like resonance between the two pair states $|27F96S\rangle$ and $|27G95P\rangle$, which are strongly dipole-coupled. A distinctive feature here is the small principal quantum number, $n = 27$, which is far away from $n' = 96$, unlike previously used detection schemes where $n \sim n'$~\cite{gunter2013observing,xu2021rydbergensembels,srakaew2023}. Consequently, relatively large EIT probe power can be used without causing too much spurious scattering of 27$F$ polaritons interacting with each other, and the read noise of the camera is easily overcome when homodyne detection is being employed. Spectroscopic measurements illustrate the working principle of the scheme and provide suitable conditions for single-shot imaging. From single-shot images of Rydberg excitations aligned in a one-dimensional chain, the pair correlation function is extracted to reveal the excitation blockade. Moreover, spatial Fourier analysis shows the presence of long-range order within the excitation chains. This work marks a major advancement of utilizing a nondestructive technique for fast and spatially-resolved imaging of Rydberg atoms. It holds particular significance in utilizing ensemble-assisted Rydberg qubits for quantum simulation and computation.

\section{F\"{o}rster-resonance-enhanced interaction}
\label{ForsterScheme}

\begin{figure*}[tbp]
\begin{center}
%\hspace*{1cm}
\includegraphics[width=\linewidth]{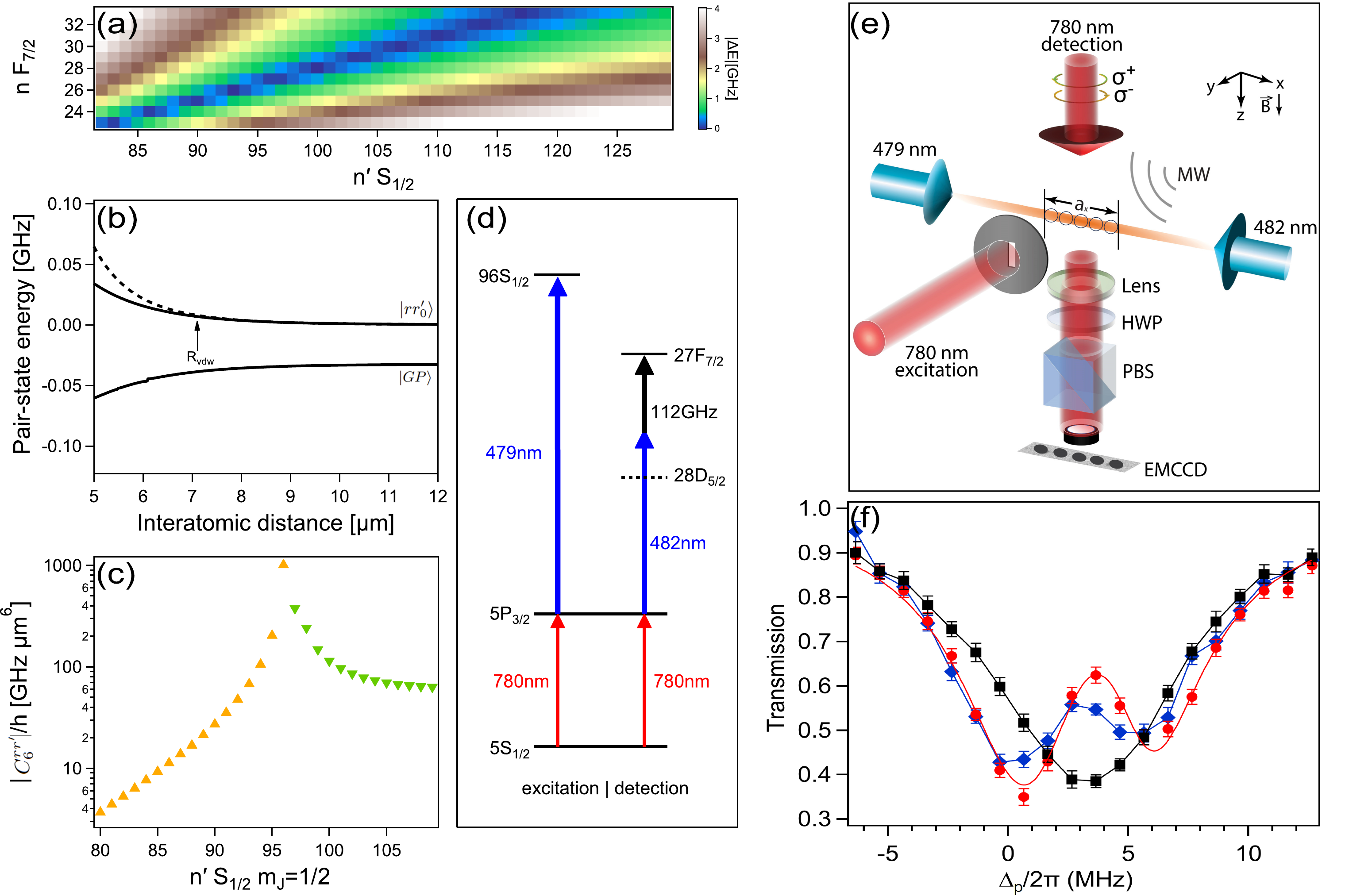}
%\hspace*{1cm}
\end{center}
\caption{\label{fig1}
(a) Pair-state energy difference $|\Delta E|$ (absolute value) for a range of $n$ and $n'$ at zero magnetic field $B = 0$. (b) Interaction potential curves of the two pair-states that asymptotically approach $|rr'_0\rangle = |27F_{7/2}, m_J = 7/2, 96S_{1/2}, m_J = 1/2\rangle$ and $|GP\rangle = |27G_{9/2}, m_J = 9/2, 95P_{3/2}, m_J = 3/2\rangle$ as $R \rightarrow \infty$ and in presence of a magnetic field $B$ = 13.3 G. The latter pair-state is the one that most strongly couples to the former. The dashed line represents the van der Waals potential $V^{rr'_0}(R)= -C_6^{rr'_0}/R^6$ obtained from fitting the upper thick solid curve for $R > 1.3 \times R_{\mathrm{vdW}}$ = 9.2 $\mu$m~\cite{vsibalic2017arc}. (c) Amplitude of the $C_6^{rr'}$ coefficients for a range of $n'$. Yellow (green) upwards (downwards) triangles correspond to negative (positive) values, indicating a repulsive (attractive) interaction. (d) Energy level schemes for Rydberg excitation and multi-photon EIT detection. (e) Experimental setup. The configurations of the atomic ensemble, the laser beams and the homodyne imaging optics are about the same as those in Ref.~\cite{du2023}. Note that the homodyne setting is not used during the spectroscopic measurements. (f) Probe light EIT spectra without prior excitation of Rydberg atoms (red circles) and with prior Rydberg excitations in states $|r'_0\rangle$ (black squares) and $|r'_1\rangle$ (blue diamonds). $\Delta_P = 0$ corresponds to the resonance of the atomic transition $|g\rangle \rightarrow |e\rangle$ in the bias magnetic field $B$ = 13.3 G. Due to the AC Stark shift caused by the off-resonance two-photon transition $|e\rangle \rightarrow |r\rangle$, the EIT peak appears around  $\Delta_P / 2\pi \simeq 3.65$ MHz. The error bars indicate the standard error of 15 measurements. The solid red line is a three-level EIT fit with fitting parameters provided in the main text. The other two solid lines provide a guide to the eye.}
\end{figure*}

The basic principle of the imaging method relies on interaction induced absorption in the vicinity of Rydberg atoms. In contrast to previously reported schemes, our approach explores an uncharted interaction resonance corresponding to the dipole-dipole-mediated energy exchange,
\begin{equation}\label{forsterresonance}
nF + n'S \Leftrightarrow nG + (n'-1)P.
\end{equation}
Here, $nF$ is the upper state of Rydberg EIT, $n'S$ is the state of the Rydberg excitations being detected, and $n$ and $n'$ are vastly different. As illustrated by the calculation in Fig.~\ref{fig1}(a), for each principal quantum number around $n \sim$ 30, there exists a specific $n' \sim$ 100 such that the pair-state energy difference $|\Delta E| = |(E_{nF} + E_{n'S}) - (E_{nG} + E_{(n'-1)P})|$ is minimal. This minimal energy difference can be as small as a few tens of MHz. For instance, at $n$ = 27, the minimal $|\Delta E| =| (E_{27F_{7/2}} + E_{96S_{1/2}}) - (E_{27G_{9/2}} + E_{95P_{3/2}})|$ is approximately 80 MHz and can be even smaller than 50 MHz for certain pair states of Zeeman sublevels in the presence of a magnetic field. The large dipole-dipole coupling described in Eq.~(\ref{forsterresonance}) can exceed the energy difference at a macroscopic distance $R_{\mathrm{vdW}}$ of a few microns, where the threshold dipole coupling is defined as $|\mu_{nFnG} \mu_{n'S(n'-1)P}|/R_{\mathrm{vdW}}^3 = |\Delta E|/2$ with $\mu_{nFnG}$ and $\mu_{n'S(n'-1)P}$ being the two dipole moments between the pair-states from the left and the right of Eq.~(\ref{forsterresonance}) respectively. Consequently, the pair interaction shift of the $|nFn'S\rangle$ state on the left side of Eq.~(\ref{forsterresonance}) is F\"{o}rster-resonance enhanced. That is, the van der Waals interaction $V(R)= -C_6/R^6$ between two atoms in these states, which is generally used to approximate the potential curve from $R = R_{\mathrm{vdW}}$ until $R \rightarrow \infty$, can possess quite a sizable $C_6$ coefficient, despite the substantial difference between $n'$ and $n$.

In our experiment, with the presence of a bias magnetic field of 13.3 G along the quantization axis $\hat{z}$, the $|r\rangle = |27F_{7/2}, m_J = 7/2\rangle$ state is used as the upper Rydberg EIT level to image Rydberg excitations in a Zeeman level $|r'_0\rangle = |96S_{1/2}, m_J = 1/2\rangle$. Shown in Fig.~\ref{fig1}(b) are the interaction potential curves of the pair states $|rr'_0\rangle$ and $|GP\rangle = |27G_{9/2}, m_J = 9/2, 95P_{3/2}, m_J = 3/2\rangle$, which is the pair state most strongly dipole-coupled to $|rr'_0\rangle$. The small $\Delta E$ = 33 MHz and strong dipole coupling ensure that the interaction potential of the pair state $|rr'_0\rangle$, $V^{rr'_0}(R)= -C_6^{rr'_0}/R^6$, possesses a large coefficient $C_6^{rr'_0}/h$ = -1010 GHz$\cdot \mu \mathrm{m}^6$. Here $C_6^{rr'_0}$ is extracted by fitting $V^{rr'_0}(R)$ to the potential curve for $R > 1.3 \times R_{vdW}$ = 9.2 $\mu$m, which asymptotically converges to the $|rr'_0\rangle$ pair state as $R \rightarrow \infty$. Plotted in Fig.~\ref{fig1}(c) is the interaction coefficient $C_6^{rr'}$, where $|r'\rangle$ = $|n'S_{1/2}, m_J = 1/2\rangle$, for a range of $n'$. A resonant peak at $n' =$ 96 can be clearly seen. In case a different detection state $n \neq 27$ is used, the resonance peak appears at a different excitation state $n'$, as implied by the plot of energy differences in Fig.~\ref{fig1}(a). This indicates that this F\"{o}rster-resonance enhanced interaction can allow state-selective imaging.

One advantage of this new scheme, utilizing an EIT detection state of low $n$ for imaging a Rydberg excitation of a high $n'$ state, is that the small interaction $V^{rr}=-C_6^{rr}/R^6$ between two atoms in the $|r\rangle$ state greatly reduces the photon scattering of the probe light~\cite{pritchard:10,han2016spectral}. As a result, this scheme allows good EIT transmission at a large enough detection probe Rabi frequency $\Omega_{p}$, which provides sufficient photon flux to surpass imaging noises under the condition of homodyne detection. Meanwhile, the interaction $V^{rr'_0}(R)= -C_6^{rr'_0}/R^6$ is quite large due to the existence of the F\"{o}rster resonance shown in Eq.~(\ref{forsterresonance}), even if the difference between $n'$ and $n$ is much larger than that typically used. These are the root causes of the increased sensitivity that warrants fast imaging of single Rydberg excitation. Another advantage of the scheme is that, compared to other works using EIT detections where $n \sim n'$ ~\cite{tiarks2014sptransistor,gorniaczyk2016,gunter2013observing,xu2021rydbergensembels,srakaew2023}, the diffusion of Rydberg excitations in $n'$ state via hopping exchanges with Rydberg populations in $n$ state does not occur, as there is no non-zero dipole moment between the $n'S$ \& ($n'$-1)$P$ states and $nF$. This is beneficial to achieve good spatial resolution.    

%
%
%%%%%%%%%%%%%%%%%%%%%%

\section{Experimental setup}
\label{experiment}

We prepare a highly elongated, cylindrical cloud of atoms in the ground state $|g\rangle = |5S_{1/2}, F = 2, m_F = 2\rangle$ with a temperature of $25 \, \mu \mathrm{K}$, which has been released from an optical dipole trap for a time of flight of 10 $\mu$s~\cite{du2023}. The cloud radially follows a Gaussian density distribution with a standard deviation of $\sigma_r = 6 \, \mu \mathrm{m}$, axially extends a couple millimeters along the $x$ direction, and has a peak density of $n_0 = 3.9\times10^{11} \, \mathrm{cm}^{-3} $ that corresponds to a mean interatomic distance of $\sim$1.37 $\mu$m.

The excitation energy level scheme is given on the left of Fig.~\ref{fig1}(d). Rydberg excitations are induced with two lasers of wavelengths 780 nm and 479 nm, on resonance with the $|g\rangle \rightarrow |e\rangle = |5P_{3/2}, F = 3, m_F = 3\rangle $  and  $|e\rangle  \rightarrow |r'_0\rangle = |96S_{1/2}, m_J = 1/2\rangle$ (or  $|e\rangle  \rightarrow |r'_1\rangle = |85S_{1/2}, m_J = 1/2\rangle$ ) transitions, respectively. While both lasers are pulsed on for a duration of 0.5 $\mu$s, the 479-nm pulse is switched on and off first, leading the 780-nm pulse by 30 ns. The configuration of the excitation beams is given in Fig.~\ref{fig1}(e). 

Due to the interaction $V^{r'_0r'_0}(R)= -C_6^{r'_0r'_0}/R^6$ between two atoms in the $|r'_0\rangle$ state, there exists an excitation blockade sphere of radius $R^{r'_0}_{b-EX} = [2 |C_6^{r'_0r'_0}|/\gamma^{r'_0}_{EIT-EX}]^{1/6}$, within which only one Rydberg excitation is allowed. Here $\gamma^{r'_0}_{EIT-EX} = \frac{(\Omega^{r'_0}_{c-EX})^2}{\Gamma}$, where $\Gamma =  2\pi \times 6.07$ MHz is the spontaneous decay rate of the intermediate state $|e\rangle$. About the same excitation blockade also arises when exciting the $|r'_1\rangle$ state. 

The parameters of two excitation beams as well as the resulting excitation Rabi frequencies are listed in Tables~\ref{ExBeamsSpatial} and~\ref{ExBeamsRabi} of Appendix B. In all the measurements of this paper, the excitation Rabi frequencies are set to give an excitation blockade radius $R^{r'}_{b-EX} = 14.5 \pm 0.2\ \mu$m. In the high density atomic cloud of our experiment, there is, with nearly unity probability, one (and only one) Rydberg excitation within a blockade sphere of radius $R^{r'}_{b-EX}$. With the thin atomic cloud in our experiment, only one Rydberg excitation is allowed radially along the $y$ and $z$ directions, and these excitations of superatoms~\cite{petrosyan2011electromagnetically,kumlin2023,shao2024} are closely packed to form an ordered array along the axial direction $x$~\cite{ates2012correlation,PetrosyanAdiabatic2013,PetrosyanCorrelations2013,vogt2017levy,du2023}. The length of the atomic cloud segment exposed to the excitation laser, $a_x$, can be adjusted, and so is the number of Rydberg excitations.

About 0.2 $\mu$s after the laser excitation, the detection pulses are turned on for 3 $\mu$s. The detection level scheme relies on multi-photon Rydberg EIT as shown on the right of Fig.~\ref{fig1}(d) with an effective two-photon coupling field driving the $|e\rangle  \rightarrow |r\rangle = |27F_{7/2}, m_F = 7/2\rangle $ transition. For spectroscopic measurements, we directly use the multi-photon Rydberg EIT of Fig.~\ref{fig1}(d) for detection in order to clearly demonstrate the spectroscopic characteristics of F\"{o}rster-resonance. However, we utilize the homodyne detection for imaging, so that individual Rydberg excitations can be resolved within 3$\mu$s exposure time. 

The homodyne detection technique has been described in detail in our previous publication~\cite{du2023}. Here is a brief summary. The weak probe light of input strength $\text{E}_{P_0}$ and a strong reference light of strength $\text{E}_R$ are the $\sigma^+$ and the $\sigma^-$ components of an elliptically polarized beam respectively, where the reference field is far detuned from the allowed weak transition $|g\rangle \rightarrow |5P_{3/2}, F = 3, m_F = 1\rangle$ and hardly interacts with atoms when passing through the cloud. At the EMCCD camera, the probe light, carrying the information about Rydberg excitations, are interfered with the strong reference light, and the overall photon count received by each camera pixel is increased enough to overcome read noise and other noises. As a result, the imaging sensitivity is sufficient to resolve individual Rydberg excitation in a 3$\mu$s exposure time. In this experiment, the intensity ratio is experimentally optimized to be $r_{R-P_0}=|\frac{\text{E}_R}{\text{E}_{P_0}}|^2 = 18$. 

Note that the recorded EIT images of the elongated atomic cloud have a 1-mm-wide field of view along the $x$ direction, much larger than $a_x$. Given in Table~\ref{detectionparameters} of Appendix B is a list of all the relevant parameters in regard to the probe beam and the fields D and M that form the effective coupling field of Rabi frequency $\Omega_{C,\text{eff}} = 2 \pi \times$5.9 MHz, in reference to the experimental configuration shown in Fig.~\ref{fig1}(e).

\section{Spectroscopic Measurements}
\label{spectroscopic}

We demonstrate the presence of the F\"{o}rster-like resonance shown in Fig.~\ref{fig1}(c) by spectroscopic measurements.
We employ EIT involving the $|r\rangle$ upper state to detect Rydberg excitations in $|r'\rangle$ states, and the relevant energy levels are illustrated in Fig.~\ref{fig1}(d). The interstate blockade radius is defined as $R_{b-r'}= [2 |C_6^{rr'}| / (\hbar \gamma_{EIT})]^{1/6}$, where $\gamma_{EIT}$ is the EIT linewidth. Inside the blockade sphere of radius $R_{b-r'}$ centered at each $|r'\rangle$ excitation, the shift of the $|r\rangle$ state due to $V^{rr'}(R)$ exceeds half of the imaging EIT linewidth $\gamma_{EIT}$ and the atoms within the sphere scatter the probe light. A larger $C_6^{rr'}$ leads to a larger $R_{b-r'}$ and consequently more absorption within the blockade sphere. Spectroscopic measurements discussed in this section show that the $|r'_0\rangle$ excitations induce more absorption than excitations in other states due to $C_6^{rr'_0}$ and the corresponding $R_{b-r'_0}$ being the largest. 

The $|r\rangle$ EIT spectra vs the frequency of the 780-nm probe light, as shown in Fig.~\ref{fig1}(f), are recorded with or without prior Rydberg excitation. The EIT spectrum with no Rydberg excitations present shows a pronounced EIT peak with up to about 50\% transmission. The solid red line is a three-level EIT fit with fitting parameters - effective optical density (OD$_{\text{eff}}$ = 1.3), effective coupling Rabi frequency ($\Omega_{C,\text{eff}} = 2 \pi \times$5.9 MHz), detuning of the coupling fields ($\Delta_{C}= 2 \pi \times$0.7 MHz), and dephasing rate of the Rydberg state ($\gamma_{gr}= 2 \pi \times$1.6 MHz). The less than 100\% transmission comes from the scattering due to the interaction $V^{rr}(R)$ between Rydberg atoms in the $|r\rangle$ state~\cite{han2016spectral}.

For acquiring EIT spectra in the presence of Rydberg impurities (Rydberg excitations amidst an atomic ensemble), we excite a one-dimensional chain of closely-packed Rydberg excitations in the $|r'\rangle$ states along the elongated cloud.  The length of the atomic cloud exposed to the excitation lasers is set to be $a_x$ = 350 $\mu$m to increase the number of Rydberg excitations for a better signal-to-noise ratio. This speeds up the measurement process, especially because we apply only the EIT detection fields here, but not homodyne detection with the reference field. By tuning the Rabi frequency of the 479-$nm$ excitation field, we keep the excitation blockade radius to be $R^{r'}_{b-EX} = 14.5 \pm 0.2\ \mu$m, which exceeds the radial size of the elongated atomic cloud $2\sigma_r$. Under such a condition, the Rydberg excitations tend to be a one-dimensional (1D) chain and closely-packed, as demonstrated and explained further in the paper. After the excitation, the detection EIT fields are turned on for spectroscopy. When the excitation state is $|r'_1 = 85S_{1/2}, m_J = 1/2\rangle$, the resulting spectrum still shows a pronounced EIT peak that is slightly reduced from that with no Rydberg excitations present. However, when the excitation state is $|r'_0= 96S_{1/2}, m_J = 1/2\rangle$, the transmission peak in this case completely disappears. This is despite the fact that the numbers of Rydberg atoms excited per unit length are about the same for exciting the $|r'_1\rangle$ and $|r'_0\rangle$ states as the same $R^{r'}_{b-EX}$ is used. The spectroscopic responses are very different because $V^{rr'_0}(R)$ is much stronger than $V^{rr'_1}(R)$ given $C_6^{rr'_1}/h$ = -9.3 GHz$\cdot \mu \mathrm{m}^6$. As a result, the interstate Rydberg-blockade sphere surrounding each Rydberg excitation has a radius of $R_{b-r'_1} = 4.0\ \mu$m vs. $R_{b-r'_0} = 8.7\ \mu$m. There are only about 80 scattering atoms inside the blockade sphere of radius $R_{b-r'_1}$ surrounding a $|r'_1\rangle$ excitation, in comparison to approximately 600 inside that of radius $R_{b-r'_0}$ surrounding a $|r'_0\rangle$ excitation. Hence, the probe transmission is much reduced in the presence of a $|r'_0\rangle$ excitation. The resulting contrast between these spectra clearly shows the effect of the F\"{o}rster resonance and illustrates the principle of this F\"{o}rster-resonance-enhanced imaging scheme.

\section{Single-shot Imaging of a Chain of Rydberg Excitations \label{images}}

To clearly demonstrate single-shot imaging of individual Rydberg excitations in the 96$S$ state, we reduce the length of the atomic cloud exposed to the excitation lasers along the symmetry axis of the atomic cloud to $a_x$ = 126 $\mu$m or $a_x$ = 17 $\mu$m. This is to reduce inhomogenous factors (e.g. atomic density along the elongated cloud), as well as to allow for a direct comparison of the cloud sections with and without excitations in single-shot images. In each experimental cycle, after exciting Rydberg atoms, we capture an image of these atoms by exposing them to probe light under EIT conditions for 3 $\mu$s, where the frequency of the probe is set to the EIT resonance position of $\Delta_P / 2\pi = 3.65$ MHz. Furthermore, we employ homodyne detection to achieve a sufficient signal-to-noise ratio for single-shot detection.

\begin{figure}[tb]
\begin{center}
\includegraphics[width=\linewidth]{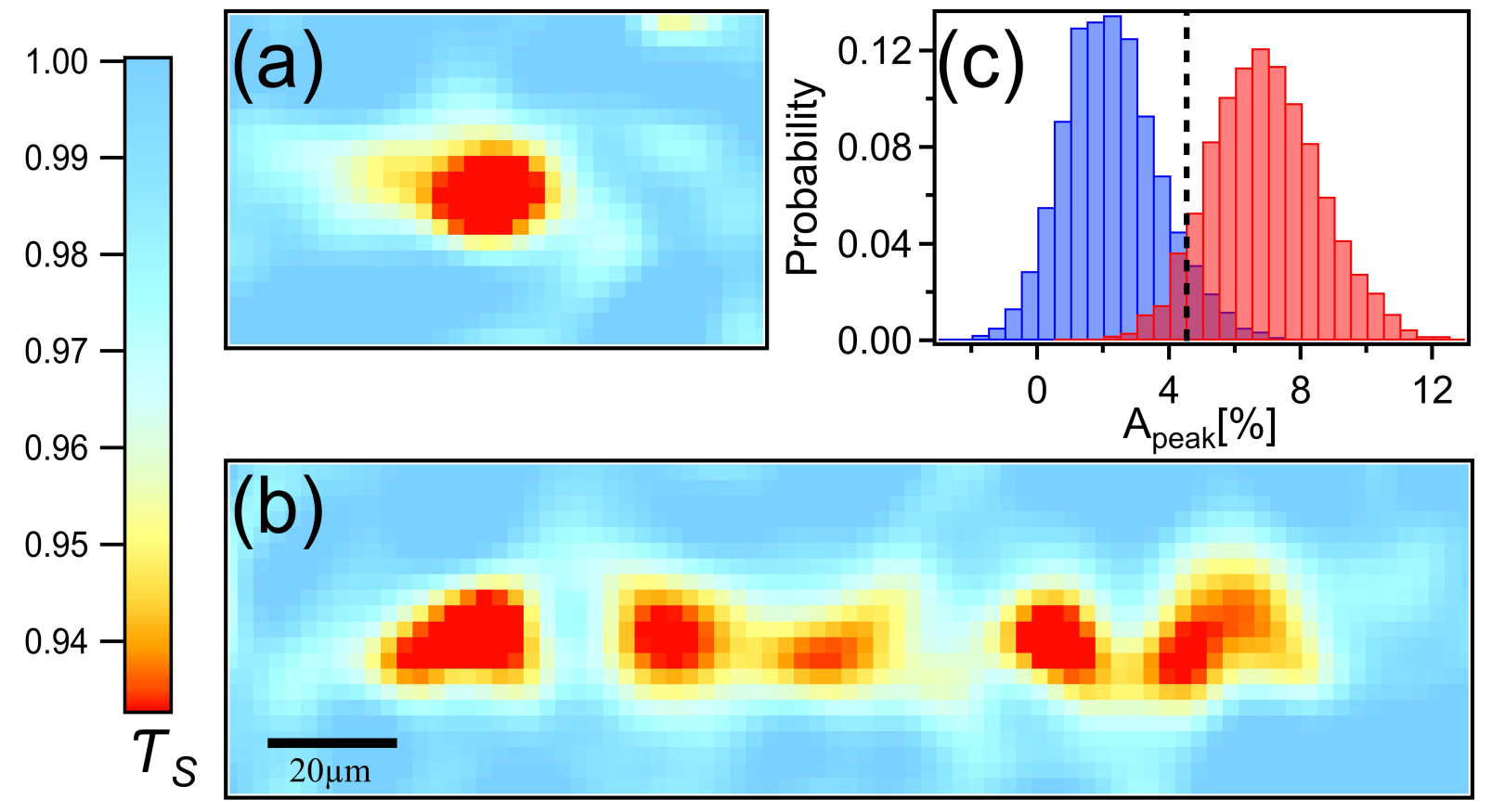}
\end{center}
\caption{\label{fig2}
(a) Gaussian smoothed (filter size = 2 pixels) single-shot image of one Rydberg excitation. (b) Gaussian smoothed (filter size = 2 pixels) single-shot image of a chain of 5 Rydberg excitations. (c) Histogram of the peak amplitudes of the absorption spots inside the excitation area (red) vs. that in an area of the same size and shape along the atomic cloud (blue). The vertical dashed line indicates the threshold $A_{thld} = 4.5\%$. The statistical distributions are based on approximately 1200 single-shot images.
}
\end{figure}

The way of processing these single-shot images of Rydberg excitations is quite similar to that used in our previous work~\cite{du2023}. After fringe removal and normalization, the obtained transmission distribution $\mc{T}(x,y)$ over each pixel position $(x,y)$ serves as the starting point for further quantitative analysis. Shown in Fig.~\ref{fig2} are smoothed images $\mc{T}_S(x,y)$ after applying Gaussian filtering to $\mc{T}(x,y)$. Pronounced absorption spots of one and five Rydberg excitations are clearly visible in Figs.~\ref{fig2}(a) and~\ref{fig2}(b), respectively, where the excitation length $a_x$ is set to be $a_x$ = 17$\ \mu$m in the former case and $a_x$ = 126 $\mu$m in the latter case. From a single-shot image as in Fig.~\ref{fig2}(b), we extract the peak amplitudes of absorption spots, defined as $A_{peak} = 1 -\mc{T}_{Smin}$, where $\mc{T}_{Smin}$ denotes local transmission minima, inside the area with Rydberg excitations. A histogram of such peak amplitudes out of many single-shot images are plotted in Fig.~\ref{fig2}(c). To make comparison, we also plot the histogram of peak amplitudes of noise spots from another area along the atomic cloud that has the same size and shape, but no Rydberg excitation. The two distributions are well separated, and can be best distinguished by the threshold peak amplitude $A_{thld}$ = 4.5\%. We evaluate the probability of identifying a Rydberg excitation in a single-shot image based on the distributions of Fig.~\ref{fig2}(c), where the true negative probability (no Rydberg atom is detected when none is present) below $A_{thld}$ is equal to the true positive probability (the presence of a Rydberg atom is detected) above $A_{thld}$~\cite{Bergschneider2018}. The fidelity of detecting a Rydberg atom is then defined as the true positive probability and estimated to be (93 $\pm$ 2)\%.

%%%%%%%%%%%%%%%%%%%%%%
%%%%%%%%%%%%%%%%%%%%
%

\begin{figure}[t]
\begin{center}
\includegraphics[width=\linewidth]{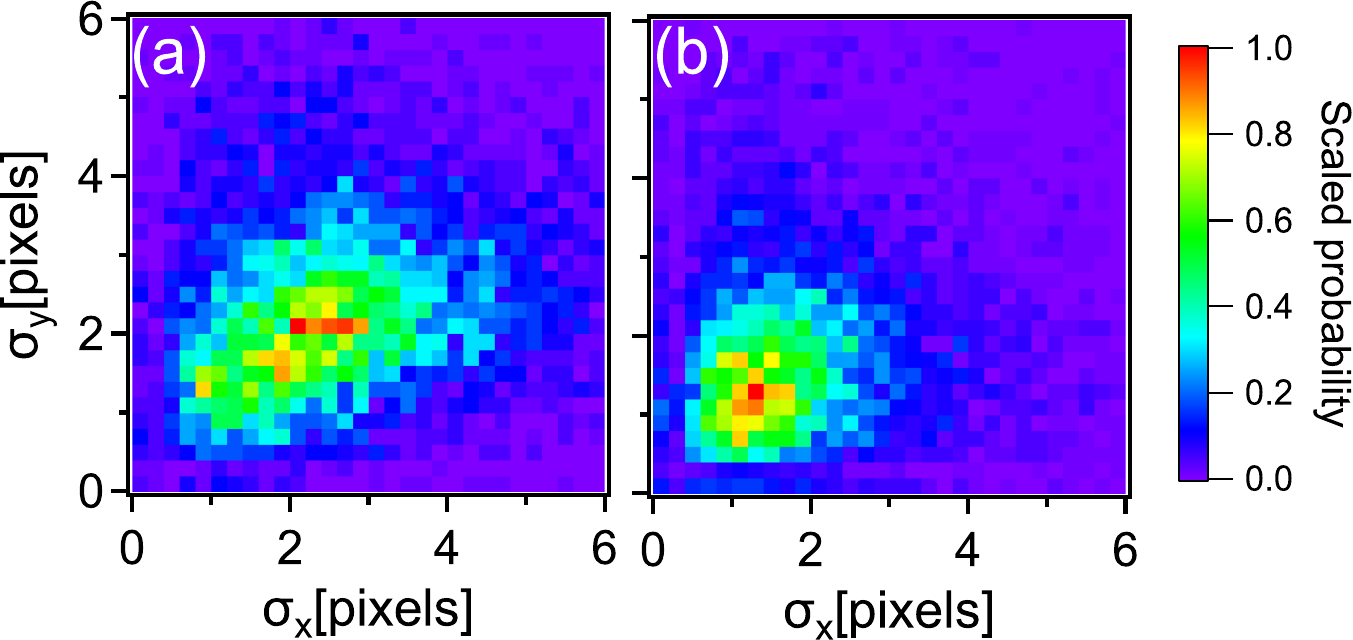}
\end{center}
\caption{\label{fig3}
Distribution of sizes $\sigma_x$ \& $\sigma_y$ extracted from (a) images of Rydberg excitations ($A_{peaks} > A_{thld}$) and (b) noise spots. The color bar indicates the normalized probability, where the value ``1.0'' corresponds to the distribution maxima in (a) and (b).
}
\end{figure}

Besides the amplitude, we also compare the sizes of the images of Rydberg excitations with those of the noise spots. We fit a 2D Gaussian distribution profile to the absorption spots in $\mc{T}(x,y)$ and extract their standard deviations $\sigma_x$ and $\sigma_y$~\cite{du2023}. As shown in Fig.~\ref{fig3}, the size distribution of the excitations is quite different from that of the noises spots. The former extends over a large range and peaks around $\sigma_x = 2.5$ pixels = 6.2 $\mu$m and $\sigma_x = 2.1$ pixels = 5.2 $\mu$m, while the latter is more localized and  peaks around $\sigma_x = \sigma_y = 1.3$ pixels = 3.2 $\mu$m. It can be seen that the two distributions are very distinct from each other. This means that in addition to the spots amplitudes, their sizes can also be used to differentiate the images of Rydberg excitations from the noise speckles. Moreover, the sizes of spots are compatible with the calculated interstate Rydberg blockade radius $R_{b-r'_0} = [2 |C_6^{rr'_0}|/ (\hbar \gamma_{EIT})]^{1/6} = 8.7\ \mu$m. \

%%%%%%%%%%%%%%%%%%%%%%%%%
%
%
%%%%%%%%%%%%%%%%%%%%%%%%%
%

\begin{figure}[tb]
\begin{center}
\includegraphics[width=\linewidth]{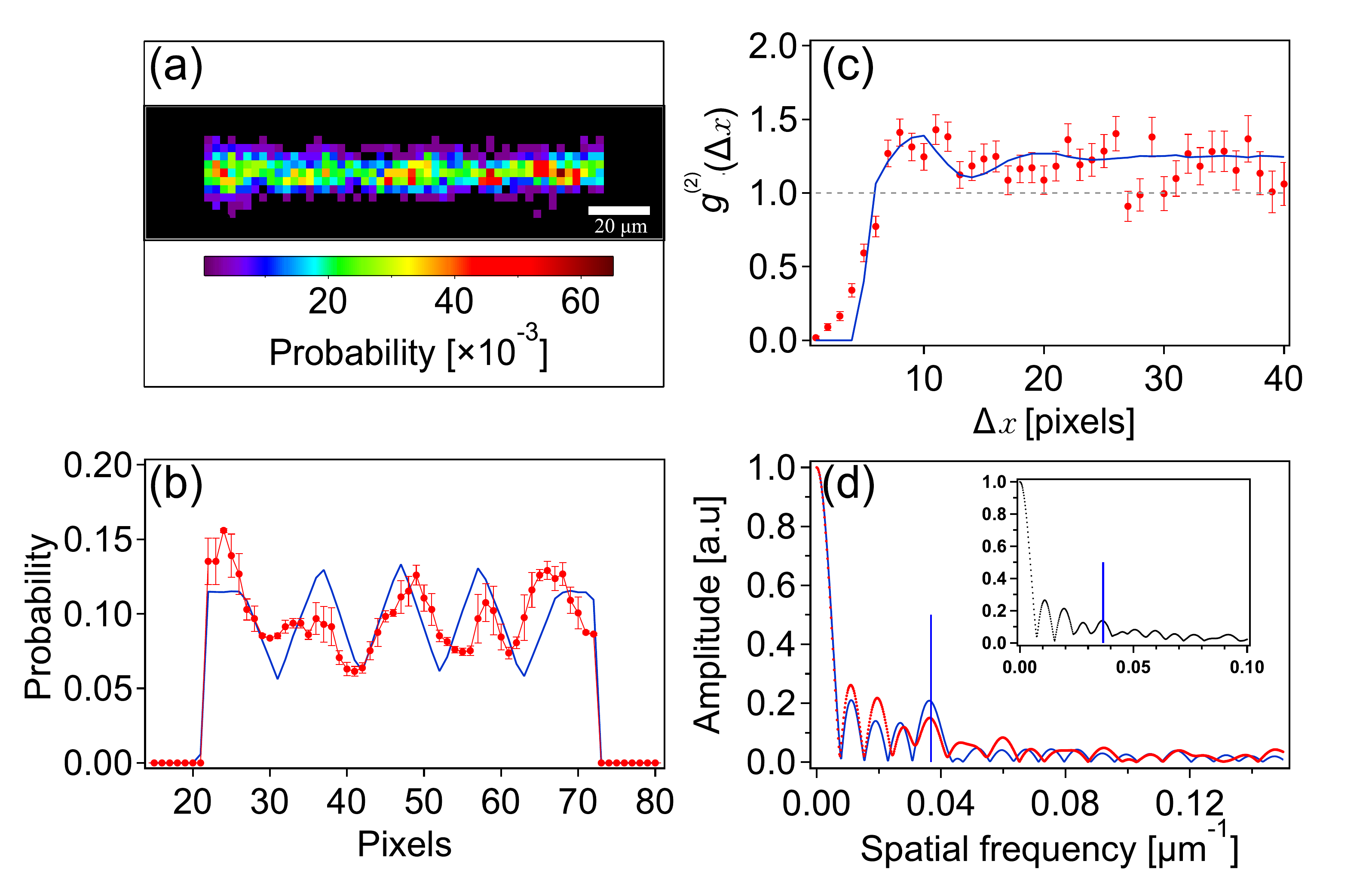}
\end{center}
\caption{\label{fig4}
(a) Two-dimensional (2D) distribution of the peak positions of 5 Rydberg excitations. This figure is obtained from single-shot images with 5 absorption spots of $A_{peak} > A_{thld}$. (b) One-dimensional distribution along the $x$ direction. The experimental data points (red markers) are derived by integrating along $y$ the 2D distribution in (a), while the blue line is obtained from the simulation of a linear chain of five Rydberg excitations. The error bars indicate the standard error of averaging three adjacent points in the 1D distribution. (c) Pair correlation function of the positions of Rydberg excitations in experimental single-shot images (red markers) and from simulated chains (blue line). The error bars represent the standard error over $\sim$ 500 images. (d) FFT spectrum of 1D spatial distributions in (b), extracted from experimental data (red markers) and from simulated distribution (blue line). Note that we have applied zero padding for computing the FFT spectrum. The inset shows the FFT spectrum of Rydberg atom 1D distributions extracted from experimental images with 4 and 6 absorption spots of $A_{peak} > A_{thld}$. The solid vertical line is the same as that in (d).
}
\end{figure}

Finally, this single-shot imaging allows us to investigate the ordered excitation of a linear Rydberg atom array in a dense atomic cloud. The excitation blockade radius $R^{r'_0}_{b-EX} = 14.7\ \mu$m leads to a chain of superatoms ~\cite{petrosyan2011electromagnetically,kumlin2023,shao2024}, corresponding to spheres of radius $R^{r'_0}_{b-EX}$ containing no more than one Rydberg atom, excited in a closely packed arrangement~\cite{PetrosyanAdiabatic2013,PetrosyanCorrelations2013}. Each imaging detection projects a Rydberg excitation into an absorption spot located within the blockade sphere of radius $R^{r'_0}_{b-EX}$. In Fig.~\ref{fig4}(a), we plot the 2D distribution of the positions of Rydberg excitations. Here the position of a Rydberg atom is taken as the location of an absorption maximum with its amplitude above the threshold $A_{peak} > A_{thld}$. Five separated distribution peaks are visible in Fig.~\ref{fig4}(a) as well as in Fig.~\ref{fig4}(b).

%
%
%
%%%%%%%%%%%%%%%%%%%%%%%%%
%
%
%%%%%%%%%%%%%%%%%%%%%%%%%
%
%

To further study the 1D spatial correlation of the Rydberg excitation chain along the $x$ direction, we compute the 1D pair correlation function as a function of $\Delta x$
\begin{equation}\label{paircorrelation}
g^{(2)}(\Delta x) = \frac{5 \Sigma_{i} \langle n_{\mathrm{ryd}}(x_i) \cdot n_{\mathrm{ryd}}(x_i + \Delta x) \rangle}{4 \Sigma_{i} \langle n_{\mathrm{ryd}}(x_i) \rangle \cdot\langle n_{\mathrm{ryd}}(x_i + \Delta x) \rangle },
\end{equation}
where $n_{\mathrm{ryd}}(x_i)$ (= 0 or 1) is the Rydberg atom detected at the pixel position $x_i$ in a single-shot image and $\langle . \rangle$ is the ensemble average over single-shot images. Note that a factor of 5/4 is included in Eq.~\eqref{paircorrelation} as there are only $N$ = 5 Rydberg excitations and this factor of $\frac{N}{N-1}$ ensures that the correlation function is equal to 1 in the uncorrelated case for a small $N$. The pair correlation function $g^{(2)}(\Delta x)$ plotted in Fig.~\ref{fig4}(c) clearly shows anti-bunching ($g^{(2)} < 1$) at small $\Delta x$, which is consistent with theory and due to the excitation blockade. Some bunching ($g^{(2)} > 1$) at large $\Delta x$ implies positive correlation of exciting Rydberg atoms. To further reveal the configuration of Rydberg excitations in the 1D chain, we perform fast Fourier transform (FFT) on the spatial distribution of the absorption spots due to Rydberg excitations. The resulting FFT spectrum is shown in Fig.~\ref{fig4}(d). A pronounced spectral peak at $1/\Delta x = 1/27.6\ \mu$m$^{-1}$ indicates the most probable separation between two adjacent Rydberg excitations. Moreover, the FFT of a simulated distribution with 5 regularly spaced peaks, as shown in Fig.~\ref{fig4}(d), agrees well with the FFT of the experimental distribution. This FFT analysis demonstrates that Rydberg atoms are excited in an ordered configuration with a nearly regular spacing of $\Delta x = 27.6\ \mu$m, which is about 2$R^{r'_0}_{b-EX}$.

Additionally, in the inset of Fig.~\ref{fig4}(d), we plot a similar FFT spectrum from experimental data of 4 and 6 Rydberg excitations being detected per image. The position of the side lobe at $1/\Delta x = 1/27.6\ \mu$m$^{-1}$ coincides with that of the spectrum from 5 regularly spaced Rydberg excitations~\cite{Note1}. This suggests that Rydberg atoms excited in such a cold-dense cloud tend to arrange in a quasi-crystalization pattern, which is generally compatible with theoretical predictions~\cite{HoningCrystallization2013,PetrosyanCorrelations2013}. Our imaging technique offers an alternative way to investigate highly-correlated Rydberg excitations~\cite{schauss2012observation}.
%
%

%Conclusion
\section{Conclusion \label{conclusion}}

In conclusion, we have achieved single-shot imaging of individual Rydberg excitations in an atomic ensemble by utilizing interaction enhanced absorption imaging in combination with homodyne detection. To reach the required signal-to-noise ratio for the detection of high-lying $n'S$ Rydberg state excitations, electromagnetically induced transparency to low $nF$ state in the vicinity of a strong F\"{o}rster resonance was chosen. We have been able to image linear chains of up to five Rydberg excitations with an imaging fidelity of 93\%. Long-range ordering was observed, suggesting onset of quasi-crystallization predicted in such systems. Further studies of time dynamics will be necessary for confirming this result.

The imaging method achieves a spatial resolution of the order of 5 $\mu$m, which is not affected by the diffusion of Rydberg excitations in $n'$ state via dipolar hopping exchanges with Rydberg populations in $n$ state~\cite{gunter2013observing,xu2021rydbergensembels,srakaew2023,Note2}. This makes it well-suited to image many-body dynamic processes. It should be noted that our fidelity of 93\%  is achieved using camera imaging instead of single photon counting in Ref.~\cite{xu2021rydbergensembels}. Compared to photon counting using APD, imaging with camera generally produces much lower fidelity due to read noise and other noises but gives quite better resolution. Only with the novel scheme in our experiment, both high fidelity and good resolution can be reached simultaneously in fast imaging detection. Furthermore, our technique can be used to detect a wide range of $n'S$ states and may be further improved by choosing lower-lying $n \ell$ states with larger $\ell$. In experiments with optical lattices and optical tweezer arrays, Rydberg excitations are detected by fluorescence imaging of ground-state atoms, which allows for spatial and temporal resolutions on the order of 1 $\mu$m and 1 ms respectively. In comparison, the imaging here has a much faster temporal resolution of 3 $\mu$s but somewhat lower spatial resolution of 5 $\mu$m, the latter of which however is comparable with typical separations of adjacent tweezer arrays. Our imaging technique could be applied for fast quantum state measurement of atomic qubits or superatom qubits~\cite{xu2021rydbergensembels} as well as \textit{in situ} probing of dipolar dynamics in many-body interacting systems~\cite{gunter2013observing}. 

\section{Acknowledgement \label{acknowlege}}
%\begin{acknowledgments}
The authors thank Klaus M{\o}lmer and Christophe Salomon for useful discussions and acknowledge the support by the National Research Foundation, Prime Ministers Office, Singapore and the Ministry of Education, Singapore under the Research Centres of Excellence programme. T. Vogt received supports from National Natural Science Foundation of China under Grant No. 12174460 and W. Li from National Natural Science Foundation of China under Grant No. 92476111.
%\end{acknowledgments}
%

\appendix

\section{Inter-state interaction potentials}
\label{FullInterPot}

%%%%%%%%%%%%%%%%%%%%%%%%%
%
\begin{figure}[tb]
\begin{center}
\includegraphics[width=0.8\linewidth]{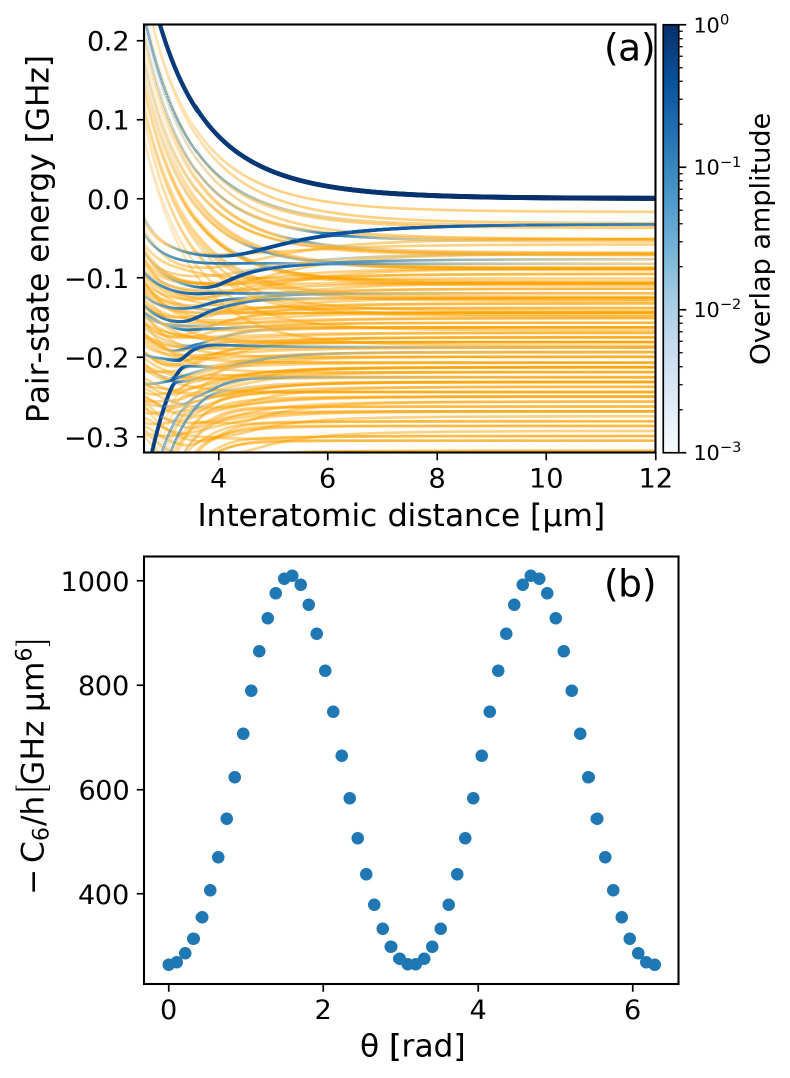}
\end{center}
\caption{\label{figS1}
(a) Calculated interaction potential curves for the pair state $|rr'_0\rangle$ at $\theta = 90^{\circ}$, where $\theta$ is the angle between the quantization axis and the internuclear axis of the pair. The color bar indicates the projection coefficient of a potential curve's corresponding eigenstate onto the unperturbed pair state $|rr'_0\rangle$. (b) The amplitude of $C^{rr'_0}_6$ coefficient plotted versus $\theta$. 
}
\end{figure}
%
%%%%%%%%%%%%%%%%%%%%%%%%%
%

We use the open-source \textit{pairinteraction} software~\cite{weber2017calculation} to calculate the interaction potential due to the coupling of  pair states near the $|rr'_0\rangle$ state, and the results are given in Fig.~\ref{figS1}. 

The non-interacting basis used for this calculation is constructed using all the pair states $|n_1,l_1,j_1,m_{j_1};n_2,l_2,j_2,m_{j_2}\rangle$ energetically near $|rr'_0\rangle$, where $\Delta_{n_1} = |n_1-27| \leq 2 \ \& \ \Delta_{l_1} = |l_1-3| \leq 2$,  $\Delta_{n_2} = |n_2-96| \leq 2 \ \&\ \ \Delta_{l_2} = |l_2-0| \leq 2$, and $j_1 \ \&\ \ m_{j_1}$ and $j_2 \ \&\ \ m_{j_2}$ take all valid values. This selection of the states results in a diagonalization space including over 2680 base vectors. These pair states are coupled together via electrical multiple interactions, the leading term of which is the dipole-dipole interaction. In Fig.~\ref{figS1}(a), the potential curves close in energy to the $|rr'_0\rangle$ state are plotted versus the interatomic distance $R$ for the case of $\theta = 90^{\circ}$. At large $R$, these energy levels asymptotically approach their unperturbed levels. At small $R$, interactions mix these states, and the energy level adiabatically evolved from the $|rr'_0\rangle$ state at $R \rightarrow \infty$ gets pushed upwards. Note that plotted in Fig. 1(b) of the main text is a simplified potential plot showing only the two most strongly dipole-coupled pair states for $R > 5\ \mu$m.

Similar to that shown in Fig.~\ref{figS1}(a), we calculate the set of potential curves for $\theta \neq 90^{\circ}$, and for each $\theta$ fit a van der Waals function $V^{rr'_0}(R)= -C^{rr'_0}_6/R^6$ to the upmost potential curve that asymptotically converges towards the $|rr'_0\rangle$ state at $R \rightarrow \infty$ (refer to Fig. 1(b) of the main text). The plot of the $C^{rr'_0}_6$ coefficients against $\theta$ is given in Fig.~\ref{figS1}(b). It can be seen that the $C^{rr'}_6$ has the largest amplitude at $\theta = 90^{\circ}$ and the smallest one at $\theta = 0^{\circ}$.

\section{Experimental parameters}
\label{experimentparameters}

Tables~\ref{ExBeamsSpatial} and~\ref{ExBeamsRabi} list the parameters of two excitation beams as well as the resulting excitation Rabi frequencies. 

Given in Table~\ref{detectionparameters} is a list of all the relevant parameters related to the detection fields, including the probe beam and the fields D and M that form the effective coupling field of Rabi frequency $\Omega_{C,\text{eff}} = 2 \pi \times$5.9 MHz, in reference to the experimental configuration shown in Fig. 1(e) of the main text.

\small
\begin{table}[!h]
\centering
\caption{\label{ExBeamsSpatial} Spatial parameters of the two excitation laser beams }
%			\begin{ruledtabular}
				\begin{tabular}{lll}
\hline \hline
			  Parameters    & 780 nm & 479 nm \\
			 			  \hline			
		  	 Propagation direction    & $-\hat{y}$  &  $\hat{x}$ \\
				Polarization     & $\hat{x}$ & $\hat{y} $   \\
                $1/e^2$ radius  & $ 3.5 $ mm (before the slits) & 31  $\mu$m \\    
\hline \hline         
			\end{tabular}
%			\end{ruledtabular}
\end{table}
\normalsize

\small
\begin{table}[!h]
\centering
\caption{\label{ExBeamsRabi} Rabi frequencies of the excitation laser beams and excitation blockade radii}  
%			\begin{ruledtabular}
			\begin{tabular}{llll}
\hline \hline
			  Excited States    & $\Omega_{P-EX} / 2 \pi$ & $\Omega_{C-EX} / 2 \pi $ & $R^{r'}_{b-EX}$ \\
			 			  \hline			
		  	   $|r'_1\rangle$  & 1.2 MHz  &  3.4 MHz & 14.4 $\mu$m \\
				  $|r'_0\rangle$ (spectroscopy)    & 2.3 MHz & 6.5 MHz & 14.7 $\mu$m  \\
				$|r'_0\rangle$ (imaging) & 2.3 MHz  & 6.5 MHz & 14.7 $\mu$m \\             
\hline \hline
			\end{tabular}
%			\end{ruledtabular}
\end{table}
\normalsize

\small
\begin{table}[tbph!]
\centering
\caption{\label{detectionparameters} Experimental parameters related to the detection fields }
%			\begin{ruledtabular}
			\begin{tabular}{ll}
\hline \hline 
			  Parameters    & Value / Unit \\
                    \hline
              of the 780 nm probe beam  &    \\
			 			  \hline			
		  	 Propagation direction    & $+\hat{z}$   \\
				Polarization     & $\hat{\sigma}^+$   \\
				$1/e^2$ radius & $ 3.4 $ mm \\
				Rabi frequency $\Omega_P$ &  $2 \pi \times 0.9\; \mathrm{MHz}$  \\
                    \hline
              of the 482 nm  beam (D field) &    \\   	
                    \hline	
                Propagation direction  & $-\hat{x}$   \\
				Polarization   & $\hat{y} $   \\
				$1/e^2$ radius & $ 46.5 \, \mu$m \\
                Frequency detuning $\Delta_D$  & $ 2 \pi \times 24 \, \mathrm{MHz} $ \\
				Rabi frequency ($\sigma^+$ component) $\Omega_D$  & $ 2 \pi \times 20.8 \, \mathrm{MHz} $ \\
                     \hline
              of the 112 GHz microwave (M field) &    \\   	
                    \hline	
                Propagation direction  & in the $\hat{x}-\hat{y}$ plane   \\
				Polarization   & in the $\hat{x}-\hat{y}$ plane   \\
		        Frequency detuning $\Delta_{M}$  & $ - 2 \pi \times  \, 26.8 \, \mathrm{MHz} $ \\
				Rabi frequency ($\sigma^+$ component) $\Omega_{M}$ & $2 \pi \times  \, 27.4\, \mathrm{MHz} $ \\
\hline \hline
			\end{tabular}

%			\end{ruledtabular}
\end{table}
\normalsize

\section{Theoretical Model}
\label{Model}

The excitation of atoms in the deep blockade regime leads to the formation of highly ordered chains of Rydberg atoms. As shown experimentally, in a majority of cases the chains contain five Rydberg atoms. Based on this finding, the model in Fig. 4 of the main text makes the simplified assumption that five Rydberg excitations are created along the axial axis of the atomic cloud, the slit area of the atomic cloud that undergoes the excitation being divided into five regions of equal length, each containing exactly one Rydberg atom. The positions of the five atoms are obtained by direct sampling. For this, the position of each atom in a given subdivision is sampled randomly, according to a uniform distribution along the axial direction, and we reset the sample of the five atoms if the blockade condition is not fulfilled. 

Monte Carlo rate equation models in Ref.~\cite{vogt2017levy} and the references therein are a more sophisticated method to simulate interacting three-level atom systems excited in the dissipative regime. However, there is no proof that these models are suitable for describing Rydberg excitations in the current experimental conditions. Moreover, the computing power needed for the Monte Carlo simulation of about 10000 atoms here is not currently within our reach. Driven dissipative systems are at present a research topic of high interest, and we believe that this work can serve as the basis for further experimental and theoretical studies.

%
%\bibliography{../bib/RydbergEIT}
%\bibliography{RydbergEIT}
%
%
%
%


%

%
%
%
%
\end{document}